\documentclass[journal]{IEEEtran}
\usepackage{multirow}
\usepackage{graphicx}
\usepackage{amssymb}
\usepackage{latexsym}
\usepackage{amsmath}
\usepackage{url}
\usepackage{hyperref}
\usepackage{cite}
\usepackage{comment}
\usepackage{subcaption}
\begin{document}
\title{Multi-Task Driven Explainable Diagnosis of COVID-19 using Chest X-ray Images}

\author{Aakarsh~Malhotra,~\IEEEmembership{Graduate Student Member,~IEEE,}
        Surbhi~Mittal,~\IEEEmembership{Graduate Student Member,~IEEE,} Puspita~Majumdar,~\IEEEmembership{Graduate Student Member,~IEEE}, Saheb~Chhabra,
        Kartik Thakral,
        Mayank Vatsa,~\IEEEmembership{Senior Member,~IEEE,}
        Richa Singh,~\IEEEmembership{Senior Member,~IEEE,}
        Santanu Chaudhury, 
        Ashwin~Pudrod,
        and Anjali~Agrawal
\IEEEcompsocitemizethanks{\IEEEcompsocthanksitem A. Malhotra, P. Majumdar, and S. Chhabra are with IIIT-Delhi, New Delhi, India 110020 (email: \{aakarshm, pushpitam, sahebc\}@iiitd.ac.in)
\IEEEcompsocthanksitem S. Mittal, K. Thakral, M. Vatsa, R. Singh, and S. Chaudhury are with IIT Jodhpur, India, 342037 (email: \{mittal.5, thakral.1, mvatsa, richa, santanuc\}@iitj.ac.in).
\IEEEcompsocthanksitem A. Pudrod is with Ashwini Hospital and Ramakant Heart Care Centre, India (email: ash147gmc@gmail.com).
\IEEEcompsocthanksitem A. Agrawal is with TeleRadiology Solutions, India 560048 (email: anjali.agrawal@telradsol.com).
}
}

\maketitle

\begin{abstract}

With increasing number of COVID-19 cases globally, all the countries are ramping up the testing numbers. While the RT-PCR kits are available in sufficient quantity in several countries, others are facing challenges with limited availability of testing kits and processing centers in remote areas. This has motivated researchers to find alternate methods of testing which are reliable, easily accessible and faster. Chest X-Ray is one of the modalities that is gaining acceptance as a screening modality. Towards this direction, the paper has two primary contributions. Firstly, we present the COVID-19 Multi-Task Network which is an automated end-to-end network for COVID-19 screening. The proposed network not only predicts whether the CXR has COVID-19 features present or not, it also performs semantic segmentation of the regions of interest to make the model explainable. Secondly, with the help of medical professionals, we manually annotate the lung regions of 9000 frontal chest radiographs taken from ChestXray-14, CheXpert and a  consolidated COVID-19 dataset. Further, 200 chest radiographs pertaining to COVID-19 patients are also annotated for semantic segmentation. This database will be released to the research community.

\end{abstract}


%
\IEEEpeerreviewmaketitle

\section{Introduction}
The COVID-19 pandemic has affected the health and well-being of people across the globe and continues its devastating effect on the global population. The total cases have increased at an alarming rate and have crossed $13$ million worldwide \cite{stats}. Increasing cases of COVID-19 patients raises the concern for effective screening of infected patients. The current process of testing for COVID-19 is time-consuming and requires availability of testing kits. This necessitates the requirement for alternative methods of screening, which is available to the general population, cost effective, time efficient, and scalable.

\begin{figure}[!t]
	\begin{center}
\begin{subfigure}{.49\textwidth}
		\centerline{\includegraphics[width=0.9\linewidth]{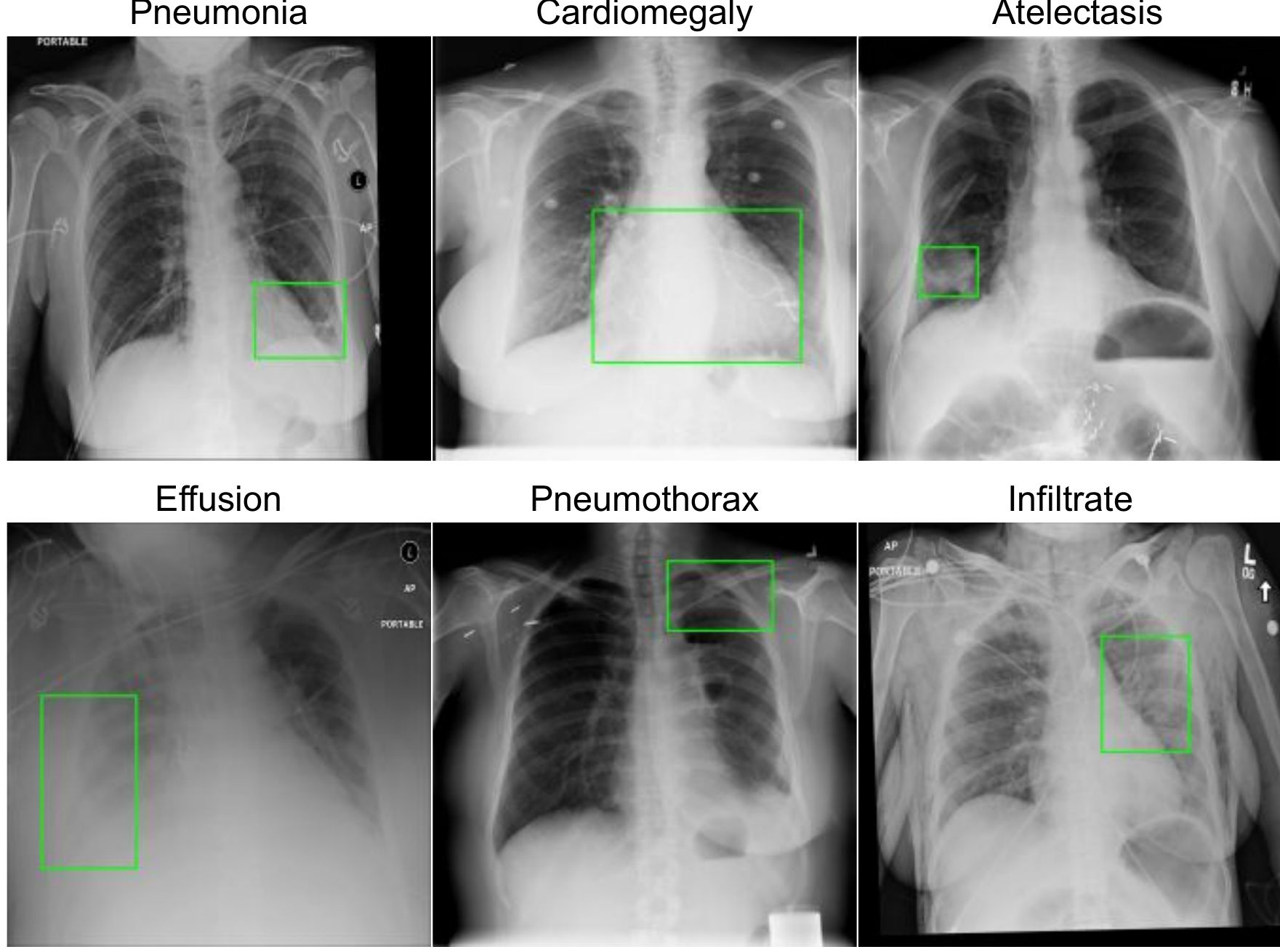}}
		\caption{}
\end{subfigure}
\newline
\vspace{3pt}
\begin{subfigure}{.49\textwidth}
				\centerline{\includegraphics[width=0.9\linewidth]{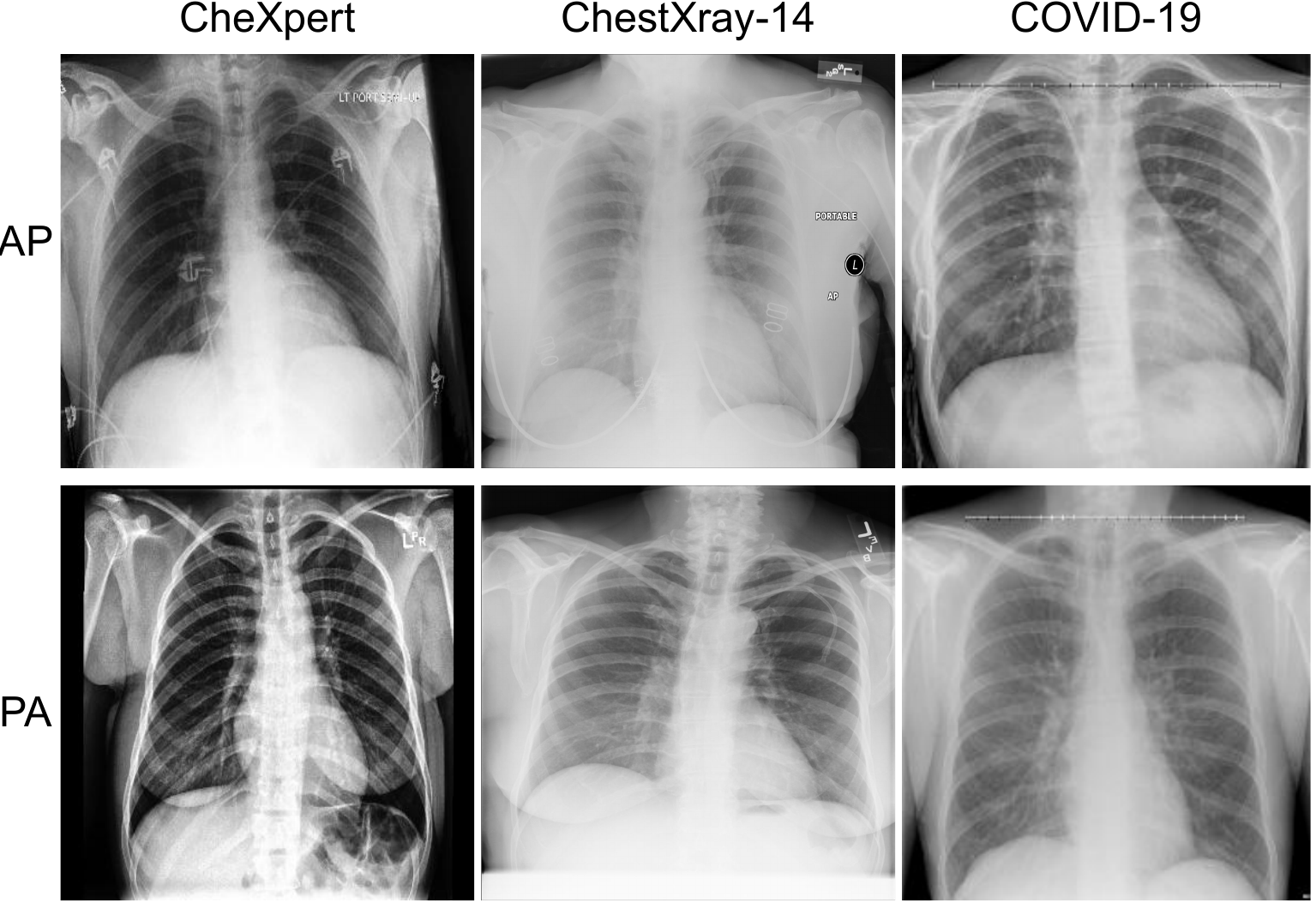}}
		\caption{}
\end{subfigure}
\end{center}
\caption{Samples of chest x-ray images. (a) Different kinds of lung abnormalities and (b) AP and PA views corresponding to CheXpert and ChestXray-14, and COVID-19 datasets. The bounding box highlights the diseased region.}\label{fig:VA}
\end{figure}

Dyspnea is a common symptom for COVID-19. Analyzing the chest X-ray, radiologists observed that it introduces specific abnormalities in a patient's lungs \cite{pan2020time}. For instance, COVID-19 pneumonia has a typical appearance on chest radiographs with bilateral peripheral patchy lung opacities, lower lung distribution, rounded morphology and absence of pleural effusion and lymphadenopathy. Fig. \ref{fig:VA} shows samples of chest x-ray images with different lung abnormalities including COVID-19. Motivated by this observation and the fact that x-ray imaging is faster, cheaper, accessible, and has scope for portability, many recent studies have proposed machine learning algorithms to predict COVID-19 using CXRs\cite{nguyen2020artificial}.

In this research, we propose a deep learning network termed as COVID-19 Multi-Task Network (CMTNet), which learns the abnormalities present in the chest x-ray images to differentiate between a COVID-19 affected lung and a Non-COVID affected lung. Since the explainability of machine learning systems, particularly for medical applications, is of paramount importance, the proposed network also incorporates the task of lung and disease segmentation. The proposed CMTNet simultaneously processes the input X-ray for semantic lung segmentation, disease localization, and healthy/unhealthy classification. Incorporating additional tasks while performing the primary task of COVID classification has multiple advantages. While processing for COVID classification, the additional segmentation tasks enforce the network to focus on lung regions and disease-affected areas only. Further, inclusion of healthy/unhealthy classification aids the CMTNet to effectively identify a healthy lung. Further, assistance from other tasks reduces dependence on enormous amounts of data required during training. The key research highlights of this work are as follows:
\begin{enumerate}
    \item Develop COVID-19 Multi-Task Network (CMTNet) for classification and segmentation of the lung and disease\footnote{In our context, the terms `abnormality', `disease', and `radiological finding' are used synonymously.} regions. The CMTNet further predicts if lungs are affected with COVID-19 or Non-COVID-19 disorders and differentiate them from healthy lungs.
    \item Inclusion of simultaneous disease segmentation in the CMTNet helps in making the decisions explainable. 
    \item Extensive evaluation and comparison against the existing deep learning algorithms for COVID-19 prediction, lung, and disease segmentation.
    \item Assemble frontal chest x-rays from various sources, that can be used for diverse tasks such as classification and semantic segmentation of lungs and disease.
    \item Creating and publicly releasing manual annotations for lung semantic segmentation for healthy, unhealthy, and COVID-19 affected X-ray images.
\end{enumerate}

\begin{figure*}[]
	\begin{center}
		\centerline{\includegraphics[width=0.85\linewidth]{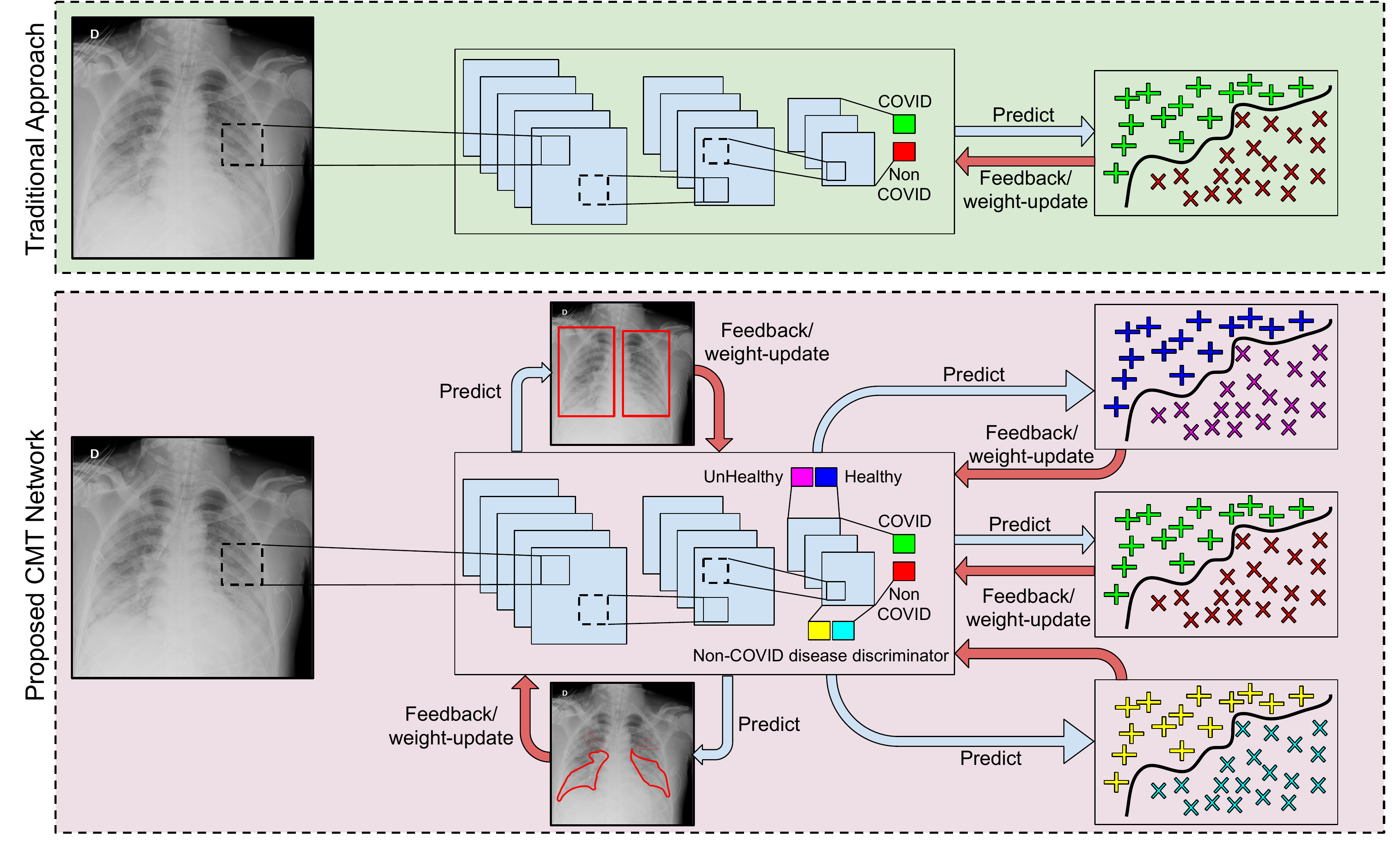}}
		\caption{The proposed CMTNet to perform multiple related tasks to improve the classification performance for COVID-19 disease diagnostics using frontal x-ray. The figure contrasts the multitask network with single task network.}
		\label{fig:Abstract}
	\end{center}
\end{figure*}

\section{Literature Review}\label{sec:lit}

Recently, researchers have proposed AI-based techniques for detecting COVID-19 using chest CT and x-ray images. Apostolopoulos and Mpesiana~\cite{apostolopoulos2020covid} explored transfer learning through various CNNs and observed that MobileNet v2~\cite{sandler2018mobilenetv2} yields the best results. Narin et al.~\cite{narin2020automatic} proposed to use three CNN models, namely, ResNet50 \cite{he2016deep}, InceptionV3 \cite{szegedy2016rethinking}, and InceptionResNetV2 \cite{szegedy2017inception} for detecting COVID-19 using chest x-ray. The authors fine-tuned these pre-trained deep models for distinguishing COVID-19 from normal x-rays and found that ResNet-50 performed the best. They used 50 chest x-ray images of COVID-19 patients from Github repository~\cite{GitHub} and 50 normal chest X-ray images~\cite{xray_pneumonia}. Abbas et al.~\cite{abbas2020classification} also performed transfer learning by fine-tuning a pre-trained AlexNet~\cite{krizhevsky2012imagenet}. Their dataset comprises 80 normal, 105 COVID-19, and 11 SARS affected lung radiographs. Cohen et al.~\cite{cohen2020predicting} predicted the severity score for COVID-19 using a deep regression model. Apart from healthy and Non-COVID disease affected frontal lung x-ray, their dataset included 94 COVID-19 affected lung x-rays (all PA views). Oh et al. \cite{oh2020deep} proposed a patch-based CNN approach for COVID-19 diagnosis. During testing, majority voting from multiple patches at different locations of lungs is performed for final decision.

For interpretation and explainability, there are limited studies. Mangal et al.\cite{mangal2020covidaid} utilized DenseNet121\cite{huang2017densely} for classification into four classes: healthy, bacterial pneumonia, viral pneumonia, and COVID-19. They showed Class Activation Maps (CAM) for interpretation. Karim et al.\cite{karim2020deepcovidexplainer} also classified into these four categories using a modified ResNet architecture. With an emphasis on explainability, the authors showed CAM and confusion matrix. On similar lines, Ghoshal and Tucker\cite{ghoshal2020estimating} showed the application of ResNet50v2\cite{he2016identity} for the above four classes. Authors interpret the results using CAM, confusion matrices, Bayesian uncertainty, and Spearman correlation.

The problem of small sample size of COVID-19 chest X-ray images was tackled by Loey et al.~\cite{loey2020within}, where they generate new COVID-19 infected images using GANs. They create a model composed of three components, namely, a backbone network, a classification head, and an anomaly detection head by Zhang et al.~\cite{zhang2020covid}. Wang and Wong~\cite{wang2020covid} introduced COVID-Net for detecting COVID-19 cases. Further, the authors investigate the predictions made by COVID-Net to gain insights on the critical factors associated with COVID-19 cases. In their work, a three-class classification is performed to distinguish COVID-19 cases from regular and Non-COVID cases.

These research demonstrate that AI-driven techniques can diagnose COVID-19 using chest x-ray images. It could potentially overcome the challenges of limited test kits and speed up the screening process of COVID-19 cases. However, a significant limitation of existing studies is that the algorithms work as a black box. These algorithms predict if the input x-ray is affected by COVID-19 or some related disease. Most studies fail to explain the decisions - for instance, which lung regions are salient for the specific decisions. Secondly, existing studies do not focus on radiological abnormalities such as consolidation, opacities, or pneumothorax. Without a clear emphasis on the lung or the abnormality, it is hard to have the explainability of an algorithm in a crucial application of COVID-19 diagnosis. Further, most of these studies work with a limited number of COVID-19 samples, with around 100 samples under most scenarios. Thirdly, as shown in Fig. \ref{fig:VA}(b), the posteroanterior (PA) and anteroposterior (AP) views of CXR images vary due to the acquisition mechanisms. While training, samples from both classes need to be considered but existing algorithms are generally silent on these details. 

\section{COVID-19 Multi-Task Network (CMTNet)}\label{sec:model}
This section provides the details of the proposed CMTNet. Multi-task networks are known to learn similar and related tasks together based on the input data. As shown in Fig. \ref{fig:Abstract}, multi-task networks have a base network with multi-objective outputs. Since each task shares the same base network, the weights are learned to be optimal for all functions jointly. The four tasks of CMTNet are (i) lung localization, (ii) disease localization, (iii) healthy/unhealthy classification and (iv) multi-label classification for COVID-19 prediction. These tasks are accomplished by using five loss functions: two for segmentation and three for classification. The details of these loss functions are described in the following subsections.  



Let $\mathbf{X}$ be the train set with $n$ images and $\mathbf{X}_i$ represent an image. $\mathbf{X}_i$ is associated with five labels, $\{\mathbf{L}_i, \mathbf{D}_i, H_i, C_i, O_i\}$ where, $\mathbf{L}_i$ and $\mathbf{D}_i$ represent the ground truth binary mask for lung and disease localization, respectively. $H_i = \{0, 1\}$, $C_i = \{0, 1\}$, and $O_i = \{0, 1\}$ represents the healthy/unhealthy, COVID/Non-COVID, and Non-COVID diseases discriminator labels, respectively. Let $f$ be the proposed CMTNet that performs the four different tasks. The task set $\mathbf{T}$ is defined as $\mathbf{T} = \{t_1, t_2, t_3, t_4\}$, where, $t_1$ and $t_2$ represent the task of lung and disease localization, respectively. $t_3$ and $t_4$ represents the task of healthy/unhealthy and COVID/Non-COVID classification, respectively.

\begin{figure*}[]
	\begin{center}
		\centerline{\includegraphics[width=1\linewidth]{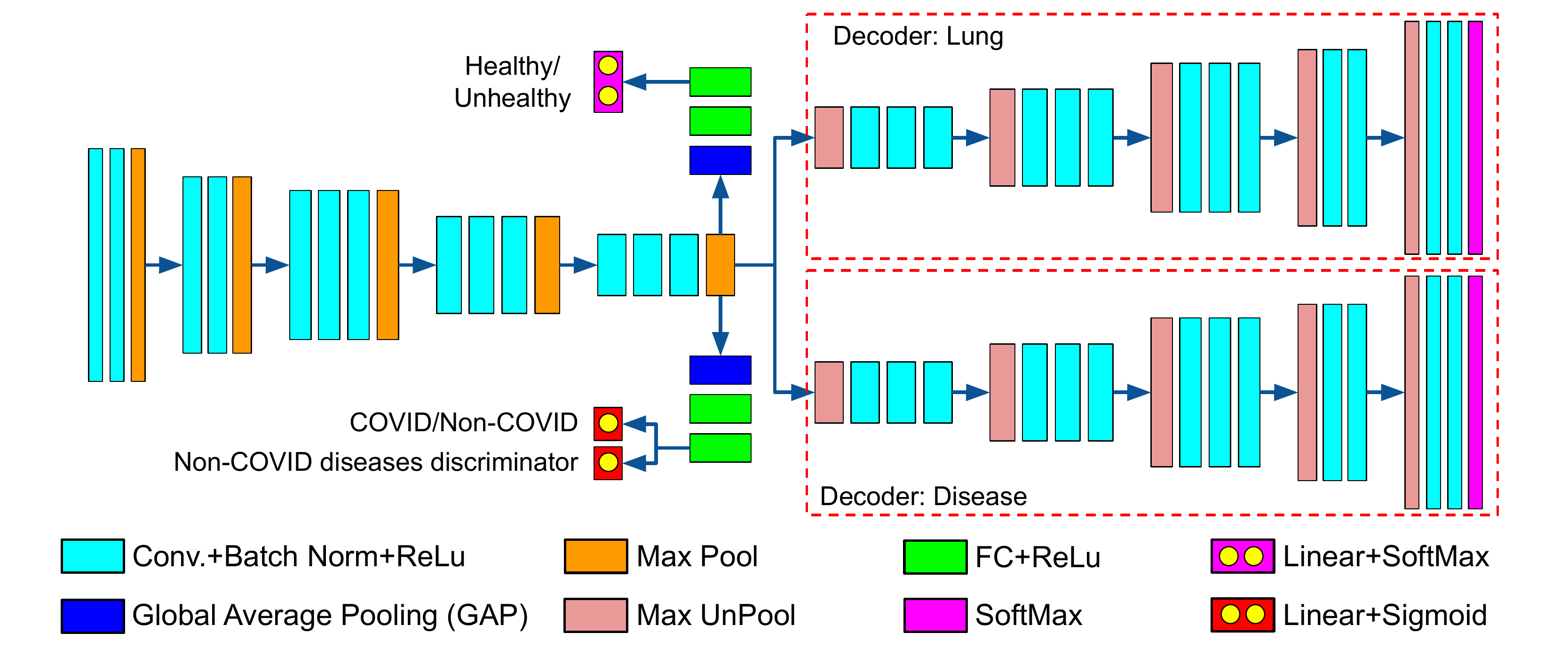}}
		\caption{Architecture of the proposed COVID-19 Multi-task Network (CMTNet), which is based on a Encoder-Decoder architecture (Best viewed in color).}
		\label{fig:model}
	\end{center}
\end{figure*}

\subsection{Segmentation Loss}

Chest x-ray of lungs contain peripheral organs along with lung regions. The primary objective of this research is to differentiate between COVID and Non-COVID samples. Since the key information lies in the lungs, the initial task is that of lung segmentation. The second segmentation loss aims to learn semantic segmentation of the diseased regions.

Lung segmentation can be achieved by learning a model that differentiates between the background and foreground lung regions. The CMTNet accomplishes this by utilizing a VGG16 Encoder-Decoder architecture~\cite{5}. The encoder has VGG16 as a base network. It has five blocks with 2, 2, 3, 3, and 3 layers of convolution + batch norm + ReLu layers, respectively. The decoder network builds upon the representation obtained from the encoder network, with a transposed architecture of the encoder network. At the final layers, the output is derived from a SoftMax layer. The output dimension equals the input spatial resolution of the X-ray image with the number of channels equaling the number of segmentation classes. Hence, the final layer consists of two channels (lung and non-lung). 

Similar to lung localization, the disease localization also builds upon the encoder representation. However, the disease localization task has a separate decoder branch and is optimized for localizing more than $20+$ lung-related disorders. For both the lung and disease localization, the gradients are backpropagated via decoder network into the encoder layers. 

Let $f_{t_1}$ and $f_{t_2}$ represent the sub-networks for lung and disease localization, respectively. For any image $\mathbf{X}_i$, the output predicted binary masks for lung and disease localization are represented as:

\begin{equation}
    \mathbf{\widehat{L}_i} = f_{t_1}(\mathbf{X}_i) \quad \text{and} \quad \mathbf{\widehat{D}_i} = f_{t_2}(\mathbf{X}_i)
\end{equation}
In this research, binary cross entropy loss is used for lung and disease localization. Mathematically, it is represented as:

\begin{multline}
    Z_{1i} = -  \sum_{x,y} \big [ \mathbf{L}_i(x,y) \log (\mathbf{\widehat{L}_i}(x,y)) + \\ (1 - \mathbf{L}_i) \log (1 - \mathbf{\widehat{L}}_i(x,y)) \big ]
\end{multline}

\begin{multline}
    Z_{2i} = -  \sum_{x,y} \big [ \mathbf{D}_i(x,y) \log (\mathbf{\widehat{D}_i}(x,y)) + \\ (1 - \mathbf{D}_i) \log (1 - \mathbf{\widehat{D}}_i(x,y)) \big ]
\end{multline}
where, $Z_{1i}$ and $Z_{2i}$ are the lung and disease loss, respectively for image $\mathbf{X}_i$. $\mathbf{L}_i(x,y)$ and $\mathbf{D}_i(x,y)$ represent the pixel value at location $(x,y)$ for lung and disease masks, respectively.

\subsection{Classification Loss}

The two classification tasks are $t_3=$ Healthy/Unhealthy classification of the lung X-ray, and $t_4=$ Multi-label classification for the presence of COVID-19 or other abnormalities. These tasks are performed using three classification loss functions. The lung and disease localization provides supervision for the three classification tasks. For healthy/unhealthy, COVID/Non-COVID, and Non-COVID diseases discrimination classification, two branches are derived over the compact encoder representation (after GAP). Each branch has three fully connected layers (FC), each followed by ReLu, ReLu, and SoftMax activation, respectively (Fig. \ref{fig:model}).


\noindent Let $f_{t_3}$ and $f_{t_4}$ represent the sub-networks for healthy/ unhealthy and multi-label classification, respectively. The output of $f_{t_3}$ for image $\mathbf{X}_i$ is represented as:

\begin{equation}
    P(H_i|\mathbf{X}_i) = f_{t_3}(\mathbf{X}_i)
\end{equation}
where, $P(H_i|\mathbf{X}_i)$ is the probability of predicting image $\mathbf{X}_i$ to $H_i$. The loss function for healthy/unhealthy classification is represented as:  

\begin{equation}
    Z_{3i} = - \sum_{H_i = \{0, 1\}} H_i \log (P(H_i|\mathbf{X}_i))
\end{equation}
where, $Z_{3i}$ represents the healthy/unhealthy loss for image $\mathbf{X}_i$. For multi-label classification, the output of sub-network $f_{t_4}$ for an image $\mathbf{X_i}$ is written as:

\begin{equation}
    [\widehat{C_i}, \widehat{O_i}] = f_{t_4}(\mathbf{X}_i)
\end{equation}

\noindent where, $\widehat{C_i}$ and $\widehat{O_i}$ represent the output predicted score ($\in$ [0, 1]) for COVID/Non-COVID and Non-COVID diseases discriminator, respectively. The radiological findings of COVID-19 pneumonia may overlap those of other viral pneumonia and acute respiratory distress syndrome due to other etiologies. The network needs supervision to segregate COVID-19 pneumonia from Non-COVID lung diseases. Hence, the joint optimization for COVID/Non-COVID along with Non-COVID diseases discrimination helps differentiate COVID-19 affected lungs from lungs affected with diseases other than COVID-19. The joint loss for predicting both COVID/Non-COVID and Non-COVID diseases discrimination is written as:

\begin{equation}
\begin{aligned}
    Z_{4i} = - \big [ C_i \log (\widehat{C}_i) + (1 - C_i) \log (1 - \widehat{C}_i) \big ] \\- \big [ O_i \log (\widehat{O}_i) + (1 - O_i) \log (1 - \widehat{O}_i) \big ]
\end{aligned}
\end{equation}

\noindent \textbf{Overall Loss Function:} It is possible that the ground truth labels or segmentation masks are not available for all the images during training. In this case, all branches of the networks will not be active during training of CMTNet. For instance, if the ground truth mask is unavailable for disease segmentation, then the sub-network $f_{t_2}$ will remain inactive and the loss $Z_{2i}$ for image $\mathbf{X}_i$ will become zero. In the same manner, other losses can have a 0/1 ``switch''. Therefore, the total loss $\mathcal{L}$ is computed as:

\begin{equation}
    \mathcal{L} = \sum_i T_{1i} Z_{1i} + T_{2i} Z_{2i} + T_{3i} Z_{3i} + T_{4i} Z_{4i}
\end{equation}
where, $T_{1i}$, $T_{2i}$, $T_{3i}$, and $T_{4i}$ are the switches pertaining to the tasks $t_1, t_2, t_3, \ \text{and} \ t_4$, respectively. The values of these switches are either 0 or 1 depending on the availability of ground truth labels/masks of the respective tasks for the $i^{th}$ image.

\begin{table}[]
\centering
\caption{Details of the databases used in the experiments.}\label{table:OtherDBsource}
\begin{tabular}{|l|c|c|c|}
\hline
\textbf{Database} & \textbf{Healthy}        & \textbf{View} & \textbf{Images} \\ \hline\hline
\multirow{4}{*}{\begin{tabular}[c]{@{}c@{}}Chest X-Ray-14 \cite{rajpurkar2017chexnet} \end{tabular}} & \multirow{2}{*}{Healthy}                                              & PA            & 4088            \\ \cline{3-4} 
                                                                         &                                                                       & AP            & 2688            \\ \cline{2-4} 
                                                                         & \multirow{2}{*}{\begin{tabular}[c]{@{}c@{}}Unhealthy\end{tabular}} & PA            & 3469            \\ \cline{3-4} 
                                                                         &                                                                       & AP            & 3115            \\ \hline
\multirow{4}{*}{\begin{tabular}[c]{@{}c@{}}CheXpert \cite{irvin2019chexpert} \end{tabular}}                                               & \multirow{2}{*}{Healthy}                                              & PA            & 1331            \\ \cline{3-4} 
                                                                         &                                                                       & AP            & 2163            \\ \cline{2-4} 
                                                                         & \multirow{2}{*}{\begin{tabular}[c]{@{}c@{}}UnHealthy\end{tabular}} & PA            & 3279            \\ \cline{3-4} 
                                                                         &                                                                       & AP            & 11305           \\ \hline
\end{tabular}
\end{table}

\begin{table}[]
\centering
\caption{Details for the COVID-19 databases used in the experiments.}\label{table:covidsource}
\begin{tabular}{|l|c|c|c|}
\hline
\textbf{Source} & \textbf{AP View} & \textbf{PA View} & \textbf{Total Images} \\\hline\hline
GitHub  \cite{GitHub}                         & 50               & 26              & 76                                                                               \\ \hline
Italy        \cite{Italy}                    & 30               & 39              & 69                                                                               \\ \hline
Spain  \cite{Spain}                          & 0                & 110             & 110                                                                              \\ \hline
RadioPaedia      \cite{RadioPaedia}              & 9                & 85              & 94                                                                               \\ \hline
BSTI    \cite{BSTI}                         & 3                & 39              & 42                                                                               \\ \hline
EuroRad     \cite{EuroRad}                     & 6                & 18              & 24                                                                               \\ \hline
\textbf{Total}                   & 98              & 317             & \textbf{415}                                                                     \\ \hline
\end{tabular}
\end{table}

\section{Experimental Details}

We next summarize the databases used for training and testing, the lung and disease annotations performed as part of this research, and the implementation details.

\subsection{Database and Protocol}\label{sec:Db}
For different tasks of the network, we require a chest X-ray database with multiple annotations and diverse properties. Thus, the database for experiments is created by combining subsets from the ChestXray-14, CheXPert, and COVID-19 infected X-ray databases. We only use frontal X-ray in our experiments from the following publicly available databases:
\begin{figure*}[]
	\begin{center}
		\centerline{\includegraphics[width=0.92\linewidth]{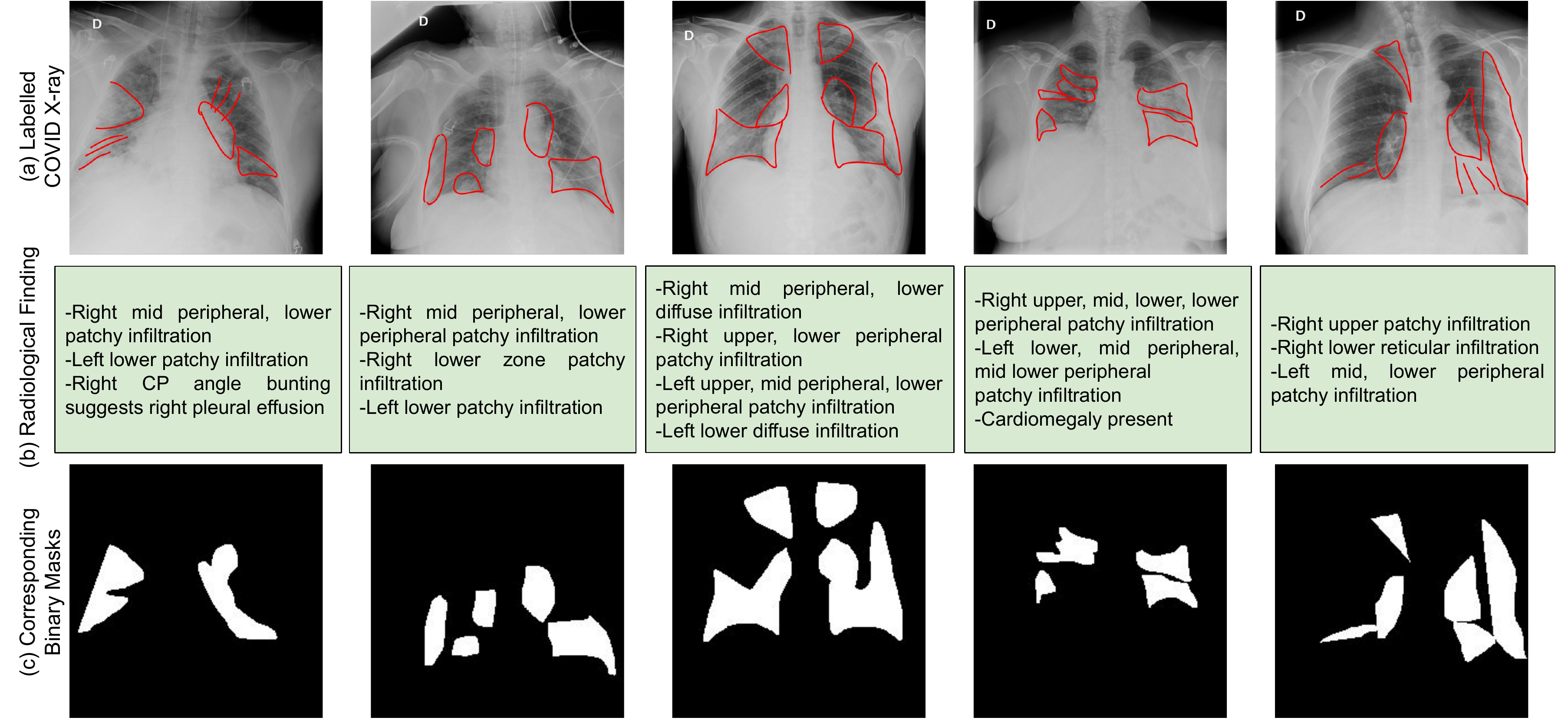}}
		\caption{Annotations provided for COVID-19 affected frontal lung x-ray images as a part of this study: (a) Labeled COVID-19 X-ray for locations of radiological finding, (b) Description of the radiological finding, (c) Corresponding binary masks for training deep semantic segmentation algorithms for disease segmentation.}
		\label{fig:GTAnnotation}
	\end{center}
\end{figure*}

\begin{table}[!t]
\centering
\caption{Details of train-test split across different parameters. Train set for Covid-19 includes augmentation.}
\label{table:dbdetails}
\begin{tabular}{|c|c|c|c|c|c|c|c|}
\hline
\multirow{2}{*}{\textbf{}}  & \multicolumn{2}{c|}{\textbf{Mask}} & \multicolumn{3}{c|}{\textbf{Disease-wise}} & \multicolumn{2}{c|}{\textbf{Views}} \\ \cline{2-8} 
                                                   & \textbf{Lung}       & \begin{tabular}[c]{@{}c@{}}\textbf{Dis-}\\ \textbf{ease}\end{tabular}      & \textbf{Normal}  & \textbf{Covid}  & \begin{tabular}[c]{@{}c@{}}\textbf{Others} \end{tabular} & \textbf{\textbf{PA}}           & \textbf{AP}          \\ \hline\hline
Train                                       & 8730       & 1456          & 8173     & 1740   & 16551         & 10161        & 16690       \\ \hline
Test                                        & 1837       & 251          & 2077     & 125    & 4097          & 2464         & 3968        \\ \hline
Total                                             & 10567      & 1707          & 10250    & 1865   & 20648         & 12625        & 20658       \\ \hline
\end{tabular}
\end{table}

\begin{itemize}
    \item \textbf{ChestXray-14 \cite{rajpurkar2017chexnet}:} 
    The dataset contains healthy and unhealthy x-ray images. It has a total of 112,120 chest x-ray images, out of which 67,310 are PA view images, and remaining 44,810 are AP view. Multiple radiographs of the same patient taken at different times are also present. From the database, we derive a subset of 13,360 images, spanning both PA and AP views. The unhealthy X-rays are labeled for one or more classes in a total of 14 classes, marking the presence of different radiological signs such as pleural effusions and consolidation. Additionally, the dataset provides localization information of abnormalities for 880 X-rays. The details of the subset drawn from ChestXray-14 is illustrated in Table \ref{table:OtherDBsource}.
   
    \item \textbf{CheXpert \cite{irvin2019chexpert}:} The CheXpert dataset contains a total of 223,414 chest x-ray images, out of which 29,420 are PA view, 161,590 are AP view, and the remaining are lateral or single lung view images. Multiple case studies of the same patient are available in the dataset. This dataset contains healthy and unhealthy X-ray images. We selected a subset of 18,078 images. Based on the radiological findings, each X-ray image is labeled positive/negative for 14 pre-defined classes (few overlapping with ChestXray-14). The details of the x-ray images selected from CheXpert database is shown in Table \ref{table:OtherDBsource}.
   
    \item \textbf{COVID-19:} For this study, we collected a total of 415 X-rays from various internet sources. The sources have a mixed number of PA and AP view frontal chest x-ray. The number of X-rays collected from each source has been summarized in Table \ref{table:covidsource}.
\end{itemize}

Since the above COVID-19 subset has limited number of images, we perform data augmentation. Each image is augmented five ways - clockwise rotation by 10$^o$, anti-clockwise rotation by 10$^o$, translation by 10 pixels in the X, Y, and XY-directions. Since pneumonia is a closely related pathology to COVID\cite{CoronaPneumonia}, we select all the pneumonia samples of the ChestXray-14 and CheXPert datasets. Further, to accommodate the variations in non-healthy x-ray samples, about 50\% more unhealthy samples are selected compared to healthy samples. AP view x-rays are prominent compared to PA views in the CheXpert dataset. Hence, we select more AP view X-ray images.

The data is split into 80\% training and 20\% testing, in a subject disjoint manner, ensuring that there is no patient overlap in the train and test sets. The details of the dataset split across different properties are specified in Table \ref{table:dbdetails}. Note that all the numbers mentioned in the table are post-augmentation.

\subsection{Lung and Disease Region Annotation}

The datasets mentioned above lack lung localization details. The proposed CMTNet requires a ground-truth lung location to identify the lung region from the x-ray. For this purpose, we manually annotated a total of about 9000 lung x-rays. These x-rays include well-balanced healthy/unhealthy, AP/PA subsets taken equally from the CheXpert and ChestXray-14 datasets. All x-ray images available for COVID-19 are also manually annotated for lung segmentation. Mask for each x-ray image has been created by drawing two solid bounding boxes, corresponding to the area covered by each lung. As a part of this study, we also plan to release the ground truth masks for the manually annotated lung regions.

The datasets included as a part of this study have only 880 disease localization annotation images (from ChestXray-14 database). For COVID-19 affected frontal lung x-ray images, we lacked disease segmentation masks. Hence, as a part of this study, the x-ray images are annotated by a radiologist for various radiological findings. The findings radiologists looked for includes: (i) atelectasis, (ii) consolidation, (iii) interstitial shadows (reticular, nodular, ground glass), (iv) pneumothorax, (v) pleural effusion, (vi) pleural thickening, (vii) cardiomegaly, and (viii) lung lesion. The experts annotated a total of 200 COVID-19 affected chest x-rays. A few sample annotations for the same can be seen in Fig. \ref{fig:GTAnnotation}(a) and the corresponding description in Fig. \ref{fig:GTAnnotation}(b). While training deep learning algorithms, the model requires binary masks as annotation. Hence, we created these masks based on the annotations (Fig. \ref{fig:GTAnnotation}(c)). We will release the ground truth binary masks to promote the training of deep semantic segmentation algorithms for abnormality localization. 

\subsection{Implementation Details}\label{sec:implementation}
The proposed Multi-task network requires input X-ray images of size 224$\times$224$\times$3. The encoder stream is initialized using a pre-trained VGG16 model. With a batch size of 16, the model is optimized over binary cross-entropy loss using Adam optimizer (learning rate $=5\times10^{-5}$). Each loss is weighted equally. The model is trained for 30 epochs on NVIDIA GeForce RTX 2080Ti and implemented in PyTorch. 

\section{Results and Analysis}
We next evaluate the performance of the proposed CMTNet for classification and localization tasks. The performance is compared with existing deep learning algorithms for COVID-19 chest radiograph studies. Further, to study the effectiveness of the proposed CMTNet, we perform experiments by selecting different combinations of sub-networks from the CMTNet. 

\begin{figure*}[!t]
	\begin{center}
		\centerline{\includegraphics[width=0.87\linewidth]{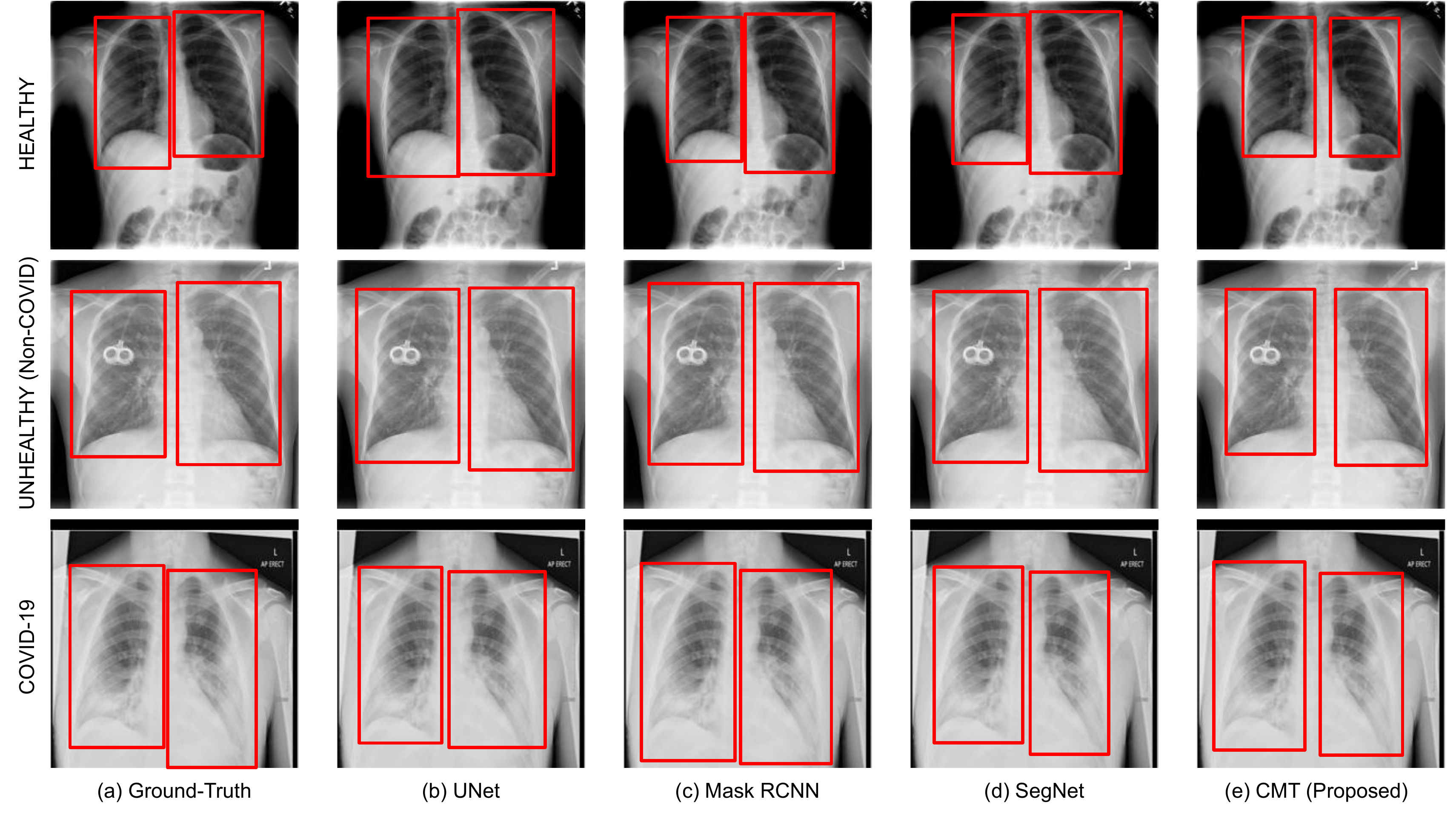}}
		\caption{Samples of lung segmentation output for existing algorithms and the proposed CMTNet.}
		\label{fig:lungSegment}
	\end{center}
\end{figure*}

\begin{figure*}[!t]
	\begin{center}
		\centerline{\includegraphics[width=0.87\linewidth]{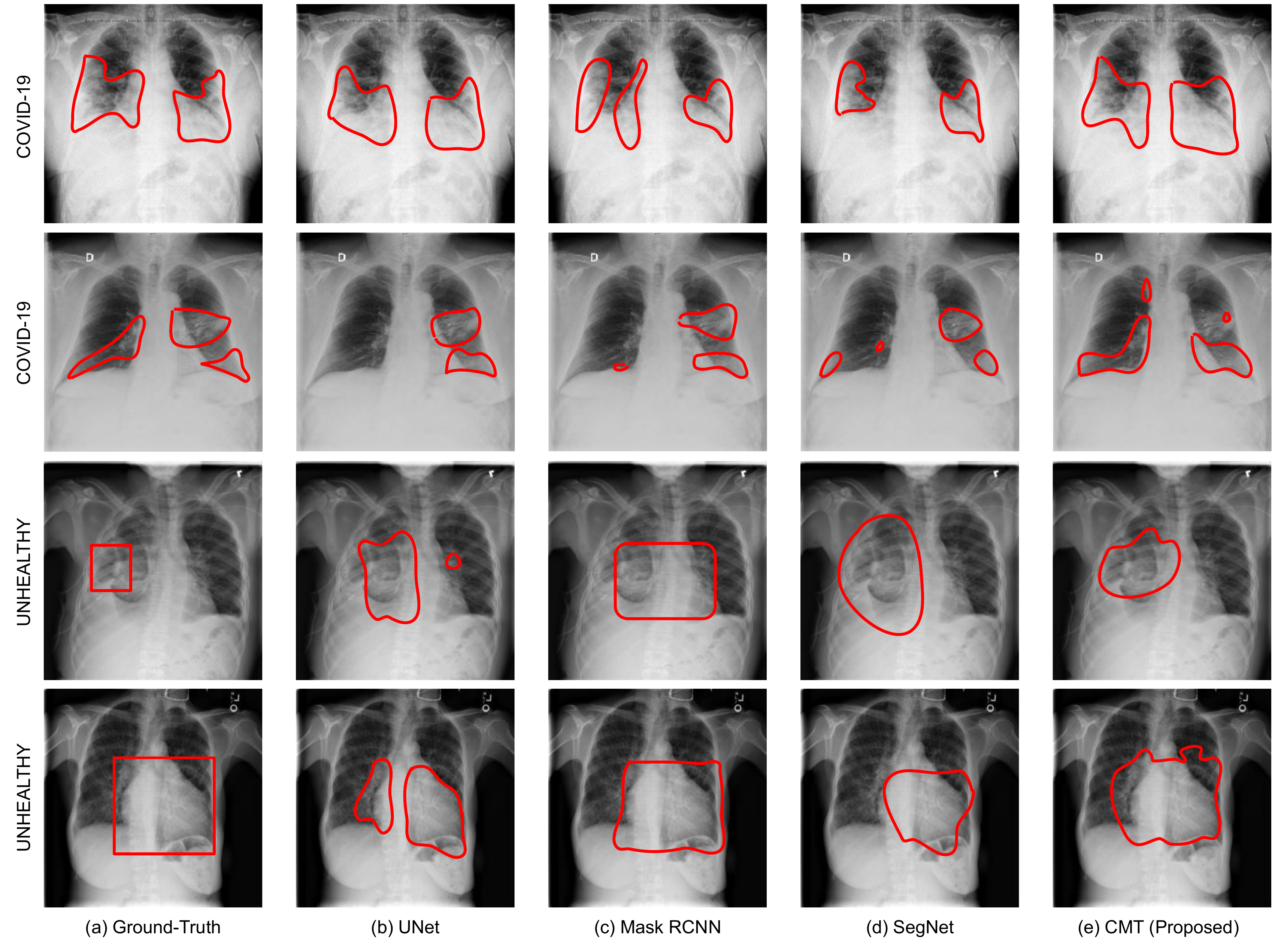}}
		\caption{Samples of semantic disease segmentation for existing algorithms and the proposed CMTNet. The x-ray images and corresponding abnormality localization for \lq\lq Unhealthy\rq\rq\textbf{ }are derived from ChestXray-14 database \cite{rajpurkar2017chexnet}.}
		\label{fig:DiseaseSegment}
	\end{center}
\end{figure*}


\subsection{Lung and Disease Localization}
In this subsection, the segmentation results of the proposed CMTNet are compared against region predictions from UNet~\cite{ronneberger2015u}, Mask RCNN~\cite{he2017mask}, and SegNet~\cite{5}. For lung segmentation, sample predictions of the proposed and existing algorithms are shown in Fig. \ref{fig:lungSegment}. Additional samples for lung segmentation can be seen in Fig. 1 of supplementary material. Inferring the sample prediction, we observe that all four algorithms perform well and give comparable results. However, the proposed CMTNet yields the most precise bound for lung segmentation. Since lung and disease localization tasks are performed simultaneously, and diseases are present within the lungs, the lung decoder network learns to focus more on the lung regions rather than the outside the lungs. 

For disease segmentation, the prediction results are shown in Fig. \ref{fig:DiseaseSegment}. Additional results are shown in Fig. 2 of the supplementary material. The first two rows of Fig. \ref{fig:DiseaseSegment} illustrate abnormalities in COVID-19 affected lungs while last two rows have abnormality localization in unhealthy but Non-COVID affected lungs. Of the four algorithms, UNet performs the worst. From the perspective of shape, Mask-RCNN tends to provide well-defined shape boundaries for Non-COVID unhealthy lungs. On the other hand, SegNet and CMTNet provide irregularly shaped predictions, localizing the radiological findings compactly. Overall, we observe that each of the four algorithms predict additional regions for the abnormalities. The detected abnormalities have false positive regions when compared to the ground-truth, sometimes localizing better than the ground-truth (for SegNet and proposed CMTNet). The same trend is elaborated in Fig. 3 of the supplementary material. 

Further, we observe that for certain abnormalities in `Unheathy' case, deep models fail to localize the abnormality. One of the reasons for this is the limited training data for abnormality localization with large variations in the diseased regions. The unhealthy Non-COVID lung abnormalities are derived from ChestXray-14, which has 700 samples corresponding to 14 labels. As a result of a small sample size for each abnormality, the networks cannot localize diseases properly. However, the proposed CMTNet has assistance from other tasks. For instance, the lung prediction task would implicitly reinforce CMTNet to predict diseases within the lung. Hence, of the four algorithms, the proposed CMTNet provides the most overlapping prediction with the ground truth.

Compared to 700 samples for 14+ different radiological findings (approx. 50 images per abnormality), the COVID-19 affected lung x-rays are 290 in number (prior to augmentation). The majority of the COVID-19 affected chest radiographs demonstrate consolidations, which tend to be bilateral and more common in lower zones\cite{covidAffectsLung}. Hence, deep models have more samples to learn the localization of COVID-19 specific abnormalities than other diseases (290 vs. 50). In retrospection, the first two rows of Fig. \ref{fig:DiseaseSegment} illustrate that all four models perform relatively better for COVID-19 localization than the last two rows of \lq\lq unhealthy\rq\rq\textit{ }localization. In most cases, each of the four models predict affected regions in the lower lung zones bilaterally. However, the proposed CMTNet outperforms other algorithms. For instance, in the first row of Fig. \ref{fig:DiseaseSegment}, both Mask-RCNN and SegNet tend to leave out the darker region in the right lung, while ground-truth and CMTNet have that region marked as diseased. Further, in the low contrast x-ray in row two, the less opaque part of the right lower lung looks darker (though being diseased). Hence, UNet fails to detect any finding in the right lower lung, while Mask RCNN and SegNet detects a few small region(s). Nevertheless, the proposed CMTNet can detect such faint differences in lung density. The same pattern can also be noticed in the low contrast x-ray (row two) of Fig. 2 in the supplementary material.


\begin{table}[]
\centering
\caption{Evaluation and comparison of the proposed CMTNet with existing learning algorithms for COVID-19 prediction. FC represents fully connected classification layers.}\label{Tab:Classification}
\begin{tabular}{|l|c|c|c|}
\hline
\multirow{2}{*}{}                                                        & \multicolumn{2}{c|}{\textbf{Sensitivity@$Y$ Specificity}}        & \multicolumn{1}{l|}{\multirow{2}{*}{\textbf{\begin{tabular}[c]{@{}l@{}}EER  (\%)\end{tabular}}}} \\ \cline{2-3}
                                                                         & \textbf{$Y$ = 99\%} & \textbf{$Y$ = 90\%} & \multicolumn{1}{l|}{}                                                                              \\ \hline\hline
DenseNet121 + FC                                                         & 60.80                       & 90.40                       & 9.82                                                                                               \\ \hline
MobileNetv2 + FC                                                         & 67.20                       & 93.60                       & 8.04                                                                                               \\ \hline
ResNet18 + FC                                                           & 56.00                       & 81.60                       & 13.78                                                                                              \\ \hline
VGG19 + FC                                                               & 50.40                       & 82.40                       & 13.70                                                                                              \\ \hline
\begin{tabular}[c]{@{}l@{}}CMTNet Embedding \\ + RDF\end{tabular}          & 79.20                       & 95.20                       & 7.34                                                                                               \\ \hline
\begin{tabular}[c]{@{}l@{}}CMTNet Embedding \\ + SVM (Sigmoid)\end{tabular} & 6.40                        & 24.00                       & 41.46                                                                                              \\ \hline
\begin{tabular}[c]{@{}l@{}}CMTNet Embedding\\ + SVM (Gaussian)\end{tabular} & 82.40                       & 88.80                       & 11.38                                                                                              \\ \hline
\begin{tabular}[c]{@{}l@{}}CMTNet Embedding \\ + SVM (RBF)\end{tabular}     & 82.40                       & 88.80                       & 11.38                                                                                              \\ \hline
\textbf{CMTNet (Proposed)}                                                          & \textbf{87.20}                       & \textbf{96.80}                       & \textbf{7.30}                                                                                               \\ \hline
\end{tabular}
\end{table}

\begin{figure}[]
	\begin{center}
		\centerline{\includegraphics[height=0.7\linewidth]{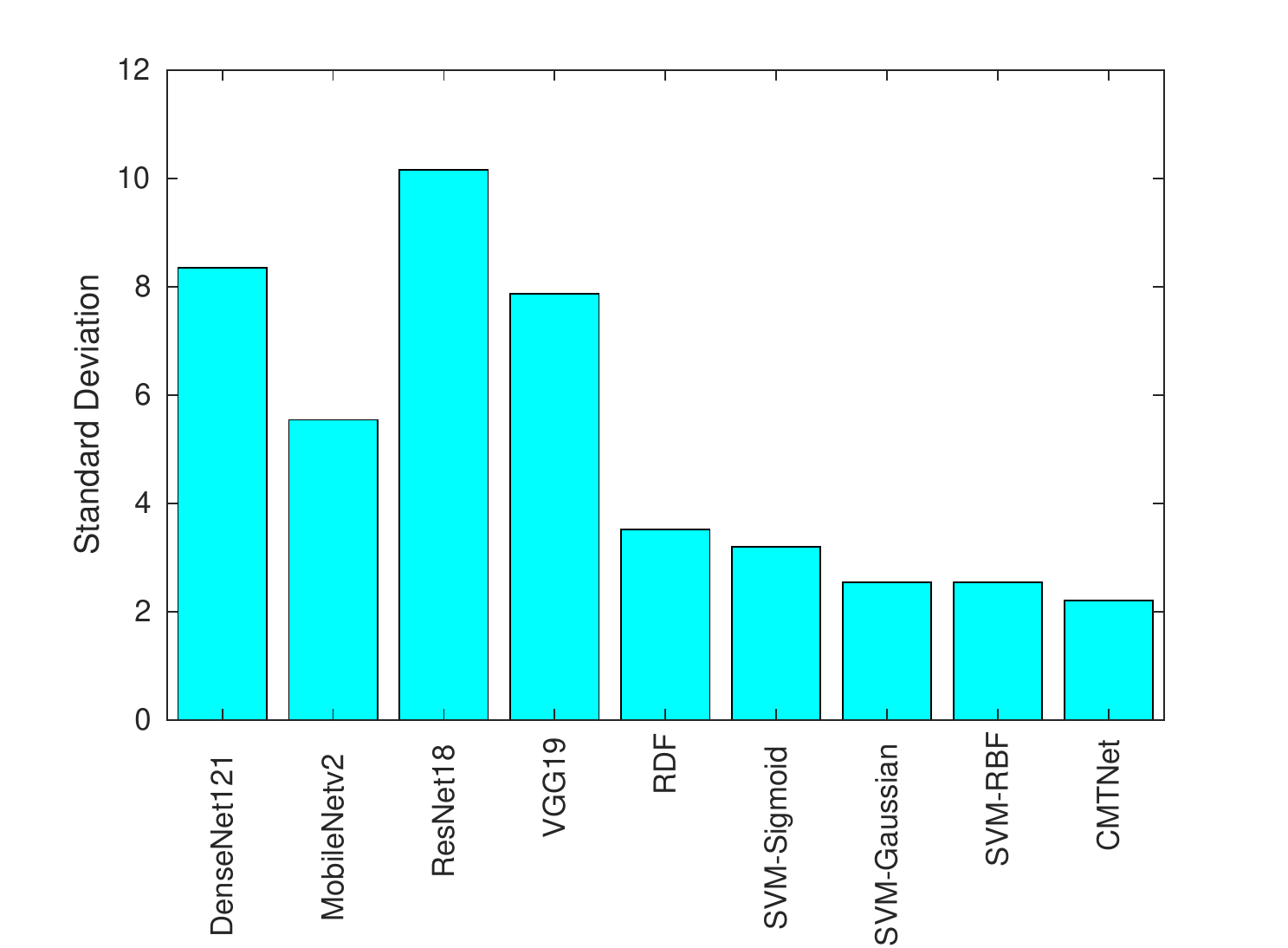}}
		\caption{Standard deviation ($\pm$) of Sensitivity (at 1\% FAR) for different algorithms. The performance is computed for different initialization of deep networks. The results show the stability in sensitivity for CMTNet, delivering consistent results for different initializations.}
		\label{fig:std}
	\end{center}
	\vspace{-15pt}
\end{figure}

\begin{figure*}[!t]
\begin{subfigure}{.5\textwidth}
  \centering
   \includegraphics[width=9cm, height=7cm]{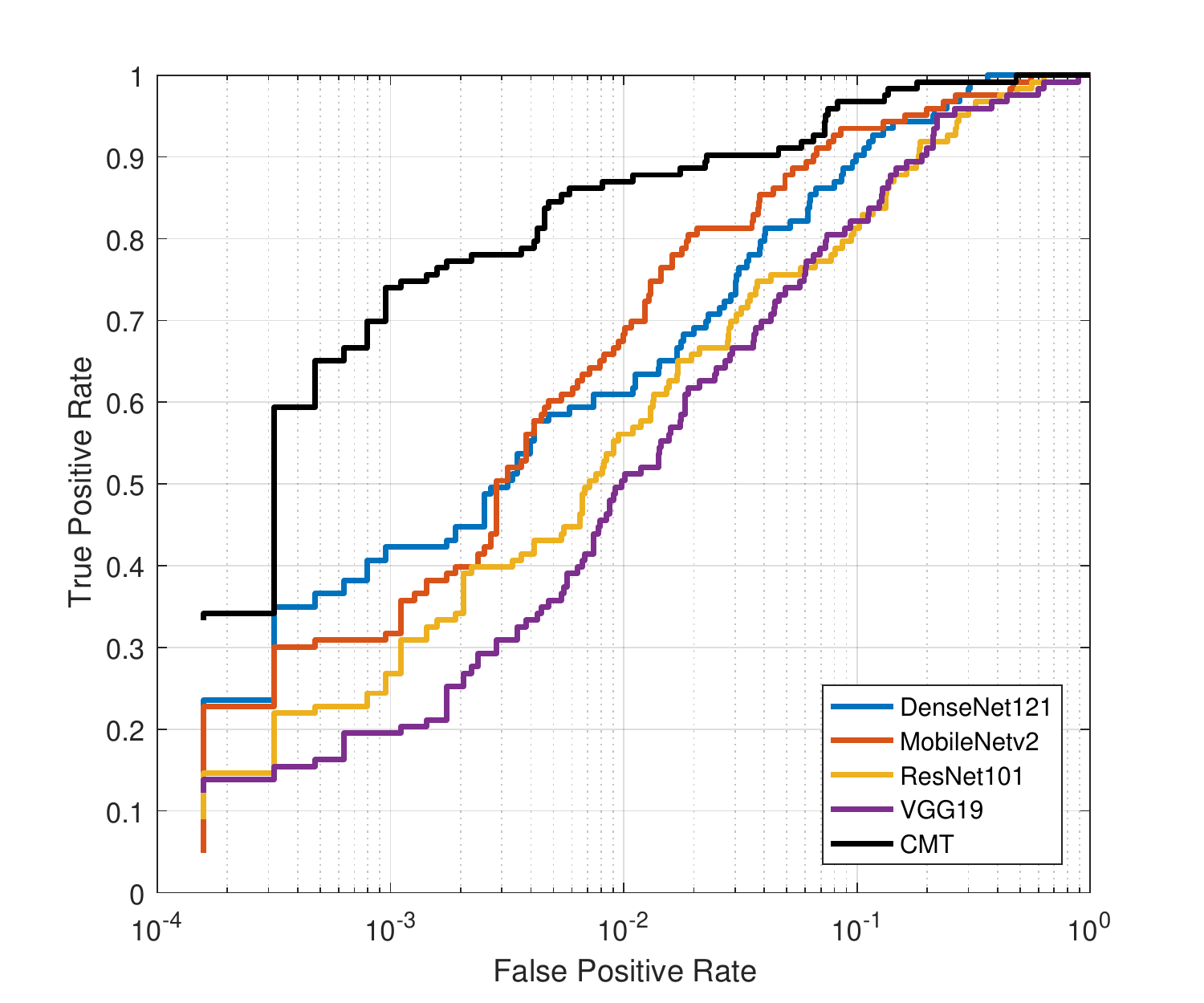}
  \caption{}
  \label{fig:sub-first}
\end{subfigure}
\begin{subfigure}{.5\textwidth}
  \centering
  \includegraphics[width=9cm, height=7cm]{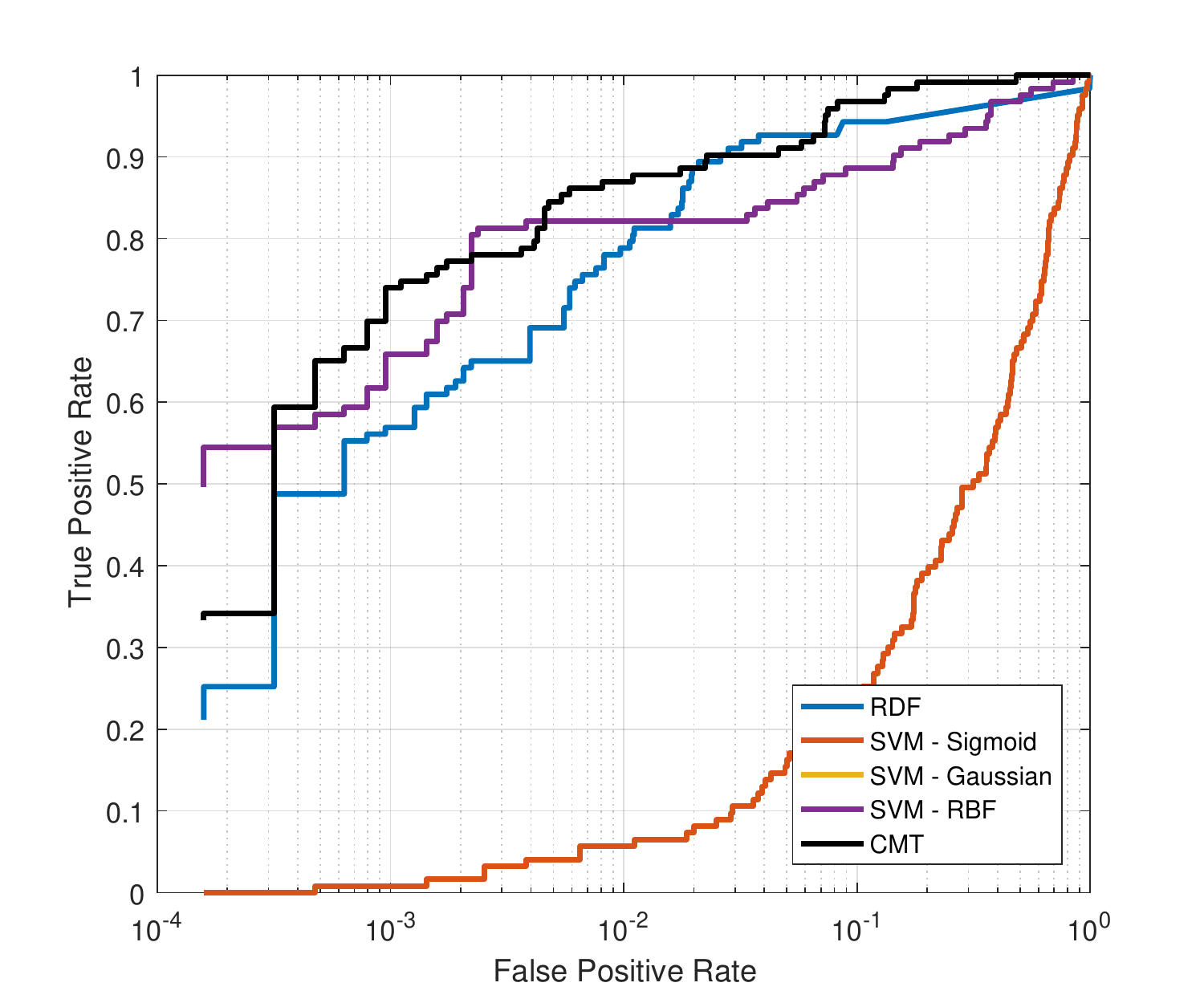}
  \caption{}
  \label{fig:sub-second}
\end{subfigure}
\caption{ROC curves summarizing the performance for COVID-19 classification: (a) comparing the proposed CMTNet with existing deep learning models, and (b) CMTNet embedding in combination with different classifiers.}
\label{fig:ROC}
\end{figure*}

\subsection{Classification}

Next, we evaluate the CMTNet's performance for healthy/unhealthy (Task 3) classification and multi-label classification of COVID-19and other diseases (Task 4). The results of the CMTNet are compared against popular deep networks. These include DenseNet121\cite{mangal2020covidaid}\cite{huang2017densely}, MobileNetv2\cite{apostolopoulos2020covid}\cite{sandler2018mobilenetv2}, ResNet18\cite{narin2020automatic}\cite{he2016deep}, and VGG19\cite{simonyan2014very}. For each of these networks, the ImageNet pre-trained version is selected. The model is then fine-tuned with the dataset and protocol used for the proposed CMTNet. Further, we draw a comparison with RDF\cite{kam1995random} and SVM\cite{suykens1999least} with three different kernels (sigmoid, gaussian, and RBF). The training of RDF and SVM is performed using feature embedding of training samples, obtained from the last encoder layer of the CMTNet.

The results for classification performance are presented in Table \ref{Tab:Classification}. It is observed that the proposed CMTNet achieves a sensitivity of 87.20\% at 99\% specificity, with an overall test classification accuracy of 98.79\%. The proposed CMTNet achieves the highest TPR and lowest EER compared to the existing algorithms. With the implicit supervision from lung and disease localization tasks, the proposed CMTNet outperforms all other existing algorithms. To show the stability of different algorithms with different initialization, the networks are three-times trained with different initialization parameters. Across different training initializations, we report the standard deviation in Sensitivity to evaluate the stability (lower standard deviation implies higher stability). As shown in Fig. \ref{fig:std}, the proposed CMTNet is the most stable algorithm across different initializations. Classifiers that use embeddings from CMTNet also report lower standard deviation. Hence, it can be inferred that CMTNet consistently provides a discriminative representation, resulting in a stable performance. Fig. \ref{fig:ROC} further shows the comparison using the ROC curves of the proposed CMTNet and existing algorithms.

\begin{figure}[]
	\begin{center}
		\centerline{\includegraphics[width=0.8\linewidth]{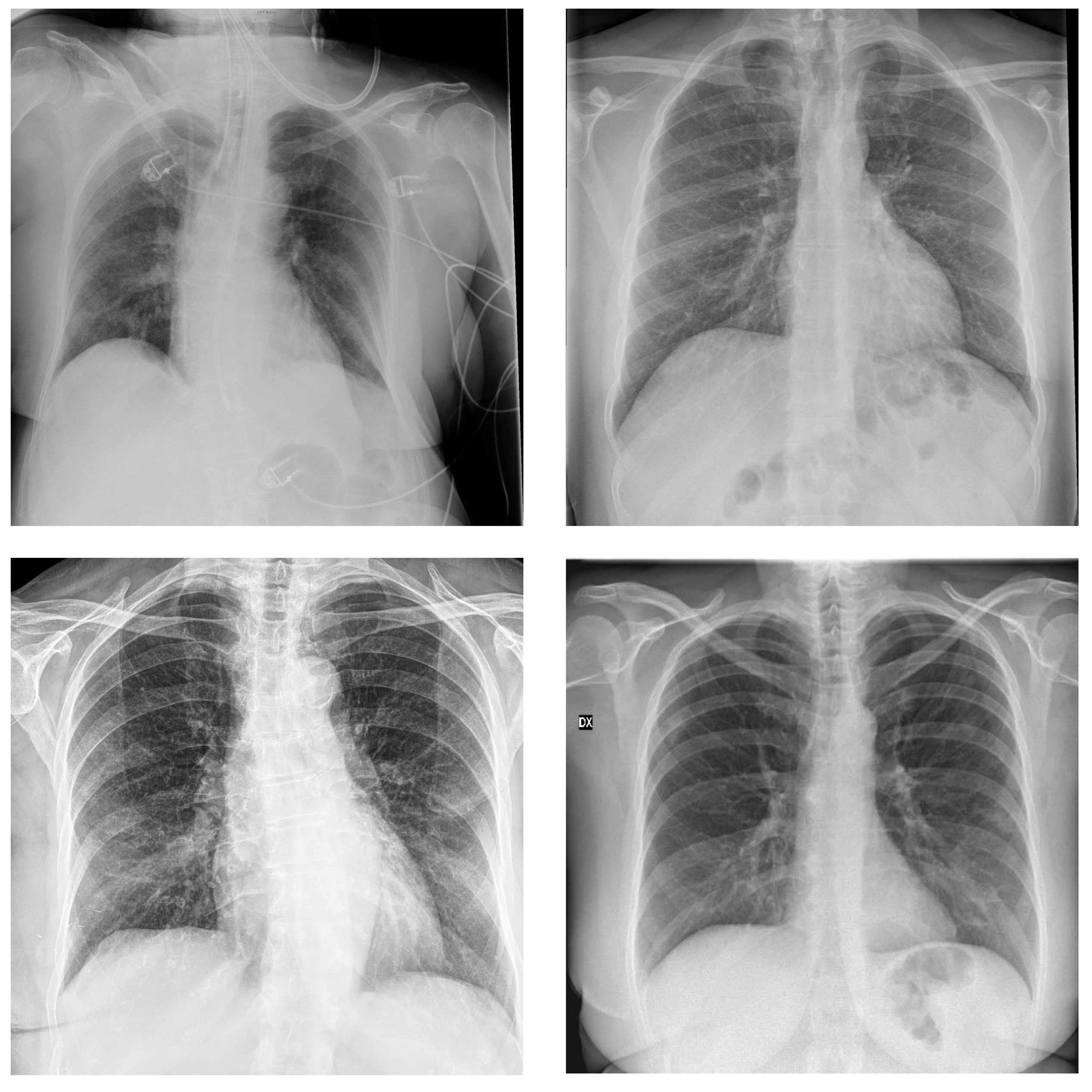}}
		\caption{COVID-19 positive case misclassified as both healthy and Non-COVID by the proposed CMTNet.}
		\label{fig:CovidTOhealthy}
	\end{center}
\end{figure}

\begin{figure}[]
	\begin{center}
		\centerline{\includegraphics[width=0.9\linewidth]{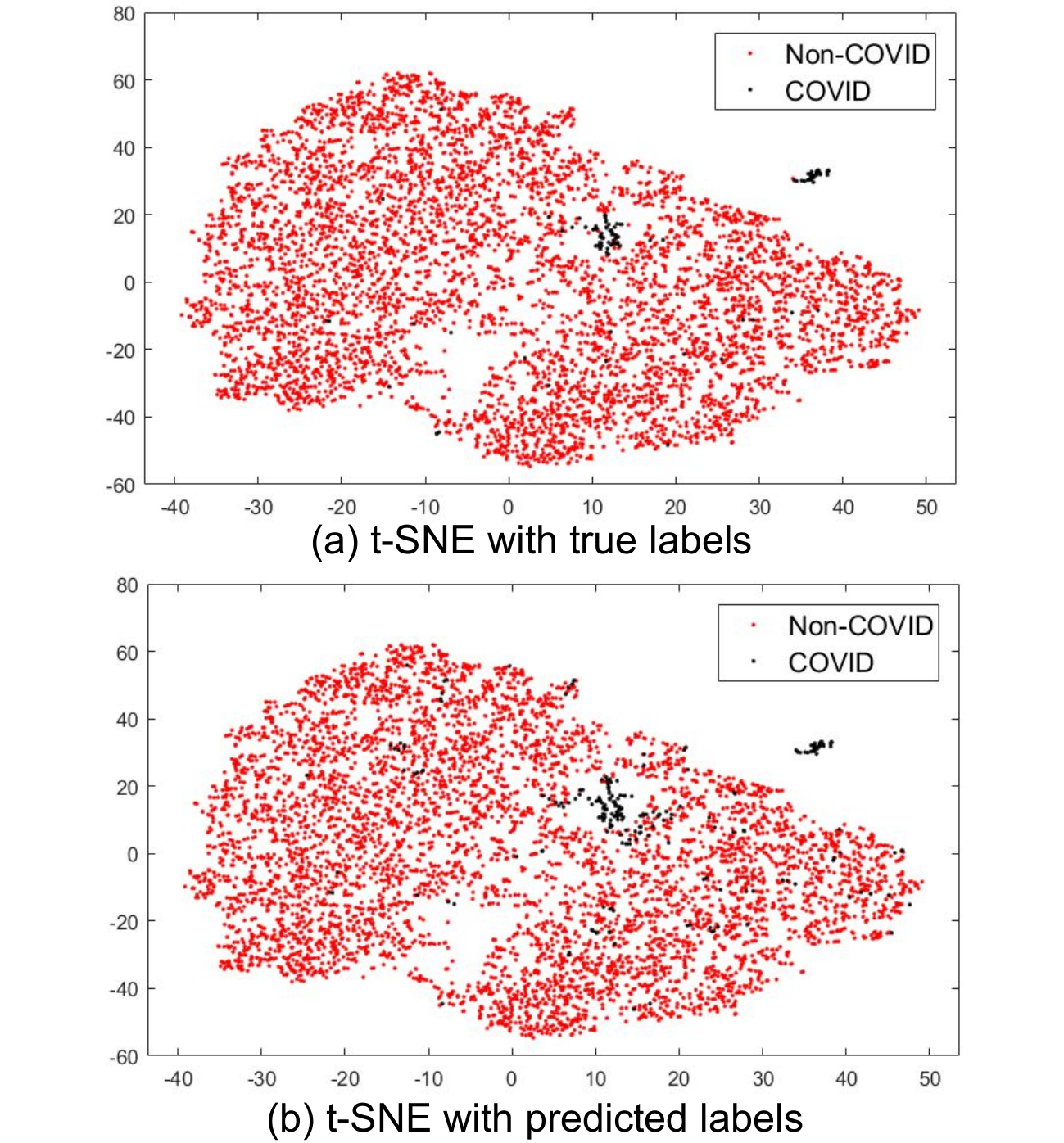}}
		\caption{Interpretation of feature representation based on (a) ground-truth and (b) predicted labels using t-SNE plot for COVID/Non-COVID classification.}
		\label{fig:tsneScore}
	\end{center}
\end{figure}

The CMTNet's classification performance for the COVID-19 samples into the healthy and unhealthy class is also analyzed. The proposed network classifies 97.25\% of COVID-19 samples into unhealthy class and 2.75\% in healthy class. The high TPR of the COVID-19 class and the majority of the COVID-19 samples being classified into unhealthy class showcase the effectiveness of the proposed network for COVID-19 detection. Overall, the classification performance of healthy/unhealthy classification is 75.17\% for all the test samples, while for Non-COVID disease classification is 73.87\%.  Based on the proposed CMTNet, Fig. \ref{fig:CovidTOhealthy} shows some of the misclassified samples where the network predicts COVID-19 positive instances (as per the RT-PCR test) into healthy (Task 3). Correspondingly, the same samples are also predicted as Non-COVID by Task 4 of the proposed CMTNet. In retrospection, we believe that minimal opacities in the lung region could be the probable cause of misclassification. This led us to check the ground truth for the hospitalization day. Of the four misclassified samples shown in Fig. \ref{fig:CovidTOhealthy}, three turned out to be the early days of the patient’s hospitalization (up to day 3). Based on these observations, it can be concluded that the CMTNet predicts an x-ray being affected when there is presence of opacities and consolidations.

\subsection{Ablation Study}
To study the importance of different tasks in the proposed CMTNet, we perform an ablation study by choosing different combinations of tasks. The four tasks in the CMTNet are Task 1: Semantic lung segmentation, Task 2: Semantic disease segmentation, Task 3: Healthy/Unhealthy classification of the lung X-ray, and Task 4: Multi-label classification for the presence of COVID-19 or other diseases. With at least one segmentation task included, we perform six different ablation study experiments, which are presented in the six rows of Table \ref{tab:ablation2}.

It is observed that for COVID-19 prediction, each task (loss function) has an important role. Removing either of the three assisting tasks deteriorates the performance. Of all these three assisting tasks, the lung segmentation task holds a pivotal role. In a COVID-19 affected x-ray, a common trait is that the lungs get affected bilaterally. Hence, a comprehensive view provided by the lung segmentation task provides more weight to lung regions, resulting in better performance with Task 1 than any other task. We perform disease segmentation and healthy/unhealthy classification since their efficacy improves in conjunction with lung segmentation and has a positive impact on the \textit{Non-COVID disease} classification prediction. As validated by the ground-truth t-SNE feature space plot (shown in Fig. \ref{fig:tsneScore}(a)), the predictions of the test COVID-19 samples (Fig. \ref{fig:tsneScore}(b)) are well separated from Non-COVID samples. It shows that the model can distinguish COVID-19 affected samples and can predict unseen test labels correctly.

\begin{table}[]
\centering
\caption{An ablation study on reducing the number of tasks and observing its effect on COVID-19 prediction.}\label{tab:ablation2}
\begin{tabular}{|c|c|}
\hline
               & \begin{tabular}[c]{@{}c@{}}COVID-19 (Sensitivity)\end{tabular} \\ \hline\hline
All 4 Tasks    & \textbf{96.80}                                                 \\ \hline
Task 1 and 4    & 94.40                                                          \\ \hline
Task 2 and 4    & 57.60                                                          \\ \hline
Task 1, 2 and 4 & 92.80                                                          \\ \hline
Task 1, 3 and 4 & 87.20                                                          \\ \hline
Task 2, 3 and 4 & 54.40                                                          \\ \hline
\end{tabular}
\end{table}

\section{Conclusion}
In the face of the SARS-CoV2 pandemic, it has become essential to perform mass screening and testing of patients. However, many countries around the world are not equipped with enough laboratory testing kits or medical personnel for the same. X-rays are amongst the most popular, cheap and widely available imaging technology across the world. This paper attempts to provide an ``explainable solution'' for detecting COVID-19 pneumonia in patients through chest radiographs. We propose the CMTNet which performs the tasks of classification and segmentation simultaneously. Experiments conducted on the different chest radiograph datasets show promising results of the proposed algorithm in COVID-19 prediction. The ablation study also supports the utilization of different tasks in the proposed multi-task network. 

\section*{Acknowledgments}
Aakarsh Malhotra is partially supported by Visvesvaraya Ph.D. Fellowship. Surbhi Mittal is partially supported by UGC-Net JRF Fellowship. Puspita Majumdar is partially supported by DST Inspire Ph.D. Fellowship. 

\ifCLASSOPTIONcaptionsoff
  \newpage
\fi



%
{\small
\bibliographystyle{IEEEtran}
\bibliography{bare_jrnl_v2}
}

\end{document}